\documentclass[aps,prl,twocolumn]{revtex4-1} 
\usepackage{amssymb,amsfonts,amsmath}
\usepackage{graphicx}   
\usepackage{epsfig}   
\usepackage{verbatim}   
\usepackage{color}      
\usepackage{hyperref}  
\usepackage{soul}
\usepackage{amsmath}
\usepackage{dcolumn}   
\usepackage{bm}        
\usepackage{amssymb}   

\begin{document}
\title{Electric-field-induced turbulent energy cascade in an oil-in-oil emulsion}
\author{Atul Varshney$^{1,2}$, Mayur Sathe$^{3}$, Shankar Ghosh$^{2}$, Anand Yethiraj$^{4}$, S. Bhattacharya$^{2}$, and J. B. Joshi$^{5}$}
\affiliation{$^1$Department of Physics of Complex Systems, Weizmann Institute of Science, Rehovot, Israel 76100\\$^2$Department of Condensed Matter Physics and Materials Science, Tata Institute of Fundamental Research, Homi Bhabha Road, Mumbai 400 005, India\\$^3$Cain Department of Chemical Engineering, Louisiana State University, LA 70803 USA\\$^4$Department of Physics and Physical Oceanography, Memorial University, St. John's, Newfoundland Labrador, Canada, A1B 3X7\\$^{5}$Homi Bhabha National Institute, Anushaktinagar, Mumbai 400 094, India}

\begin{abstract}
We observe electro-hydrodynamically driven turbulent flows at low Reynolds numbers in a two-fluid emulsion consisting of micron-scale droplets. In the presence of electric fields, the droplets produce interacting hydrodynamic flows which result in a dynamical organization at a spatial scale much larger than the size of the individual droplets. We characterize the dynamics associated with these structures by both video imaging and a simultaneous, \emph{in situ}, measurement of the time variation of the bulk Reynolds stress with a rheometer. The results display scale invariance in the energy spectra in both space and time.
\end{abstract}

\maketitle
Turbulence is represented as a cascade of energy which flows from larger scale eddies to smaller ones where it is dissipated \cite{Tennekes_1972} due to viscosity. For Newtonian fluids, in pipe flow driven by pressure gradients, the flow becomes unsteady when nonlinear inertial forces exceed the dissipative viscous forces which happens at Reynolds numbers $Re \sim 2000$ \cite{Reynolds_1883, Avila_2011}, where $Re$ is the ratio of inertial to viscous forces. Turbulence has also been observed at a very low $Re$ in non-Newtonian fluids, e.g. in polymer solutions \cite{Groisman_2000}, and in dense suspensions of ``active'' self-driven systems \cite{Wensink_2012, Dunkel_2013}. The source of the nonlinearity in polymer solutions has been attributed to the presence of elastic stresses that arise from the conformational change of polymer molecules during the flow, whereas, in active systems the non-linearity arises from mutual interactions. 
In this Letter, we present a model system with meso-scale components $-$ a micron-scale oil-in-oil emulsion $-$ where we observe the unsteady flows characteristic of turbulence. We demonstrate two levels of control: we control the strength of the hydrodynamic interactions with an external electric field, and the energy-injection lengthscale via the size of the droplets. We obtain multiple measures of turbulence using optical microscopy and rheometry concurrently, and observe self-consistent power law scalings of both spatial and temporal energy spectra.

In the presence of an electric field, the difference in the conductivity of a liquid drop with its surrounding fluid results in electrohydrodynamic forces, first reported by Melcher and Taylor \cite{melcher} and studied extensively thereafter \cite{saville,salipante}. 
When many proximate droplets respond to electrohydrodynamic forces, the local asymmetry results in both translational and rotational motion on the droplets. Electric field experiments describing the frequency-dependent phase diagram in an oil-in-oil emulsion where many droplets can interact with each other was reported by us earlier \cite{Varshney}, and dc fields were found to provide the strongest hydrodynamic interactions.  Our system consists of an oil-in-oil emulsion,  prepared by mixing silicone oil (a dielectric) and castor oil (a leaky dielectric) in a volume ratio $1:3$. The materials constants for silicone and castor oils are tabulated in Supplementary Table 1. 
\begin{figure}[hbt]
\includegraphics[scale =0.32]{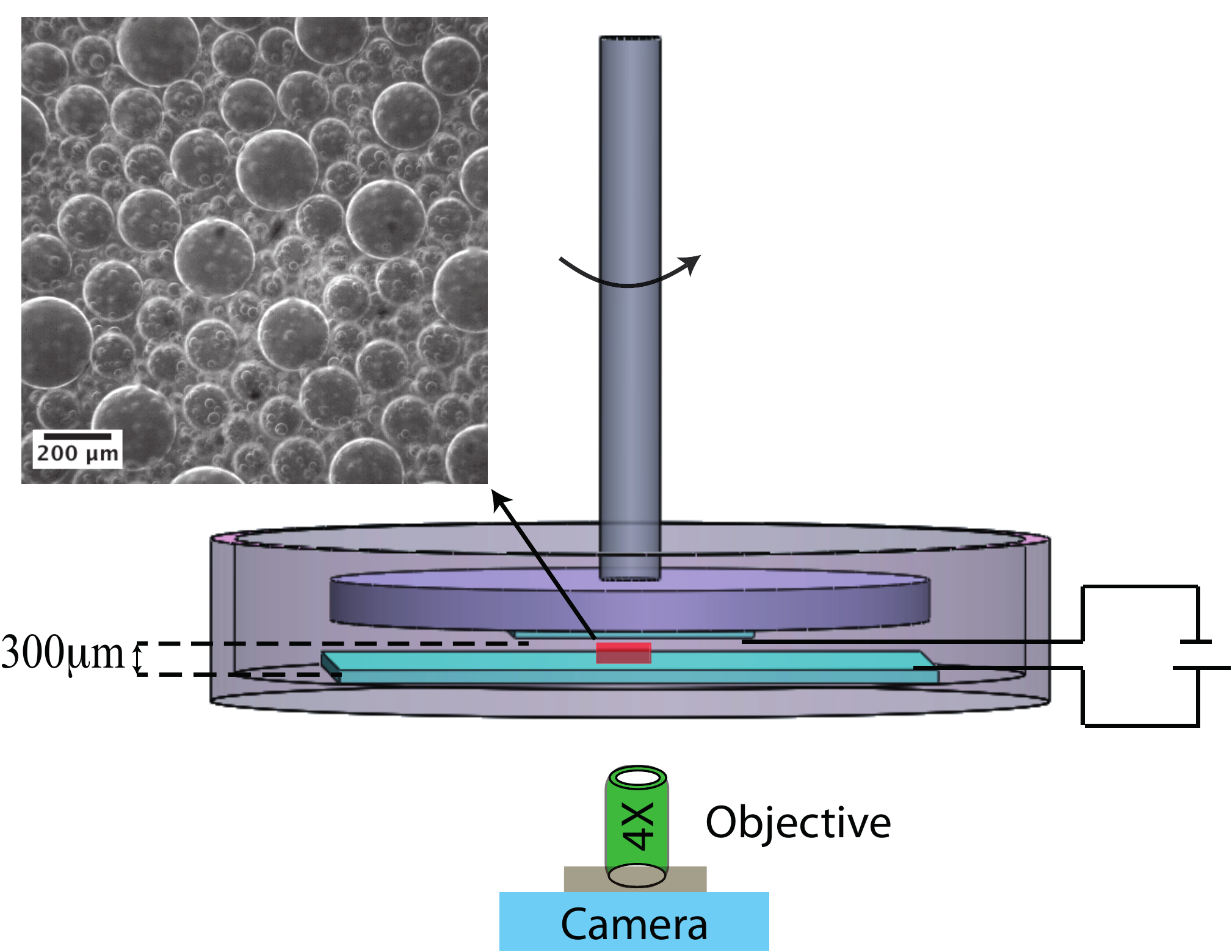}
\caption{A rheometer, shown schematically, is used to measure shear stress in an oil-oil emulsion system. Two ITO electrodes (cyan) are glued onto both the rotating top plate as well as the bottom surface of the stationary cylindrical cup of the rheometer. A dc electric field is applied across the electrodes. The system is imaged from the bottom using a high-speed camera (PCO 1200s) with a 4X Nikon Plan Apo microscope objective (NA = 0.2, working distance = 15.7mm). The illumination is provided from the top by a customized LED light source. The micrograph shown (top left) is a snapshot of the emulsion at zero field. The red rectangular box shows the region that is being imaged between the electrodes. }
\label{figs1_turb}
\end{figure}
In the absence of an external field, the oil-in-oil mixture is Newtonian, exhibiting a clean, linear dependence of stress on strain rate.  
The experimental set-up is shown schematically  in Fig. \ref{figs1_turb}.
The emulsion was placed between two ITO coated electrodes (spaced $300~\mu m$ apart), mounted in the parallel plate geometry onto a rheometer (Anton Paar; MCR301). An external dc electric field is applied  across the electrodes by using a high voltage source (SRS PS375).  
Upon increasing the electric field from $E= 0$ to $6.7~V/\mu m$, we observe a violent breakup of large drops into smaller droplets (Supplementary Movie 1), coupled with unsteady droplet flows. The droplets continually split into daughter structures and recombine: while the interfacial electrostatic stresses aid their break-up, surface tension provides the stabilizing force opposing the production of smaller sized drops. At each electric field, we observe a steady-state distribution of droplet sizes $N(a)$, which is shown as a function of the drop radius $a$ for  $E>6~V/\mu m$ in Fig. \ref{fig5_turb}.  For fields larger than $E = 6.7~V/\mu m$ there is a steady state droplet size distribution, and a log-normal form (with a mean droplet radius $=22\pm1\mu m$) provides a good fit to the size distributions, as indicated by Perlekar and coworkers \cite{Prasad}. We identify the droplets themselves to be the physical objects into which the energy is injected.
\begin{figure}[hbt]
\includegraphics[scale=0.3]{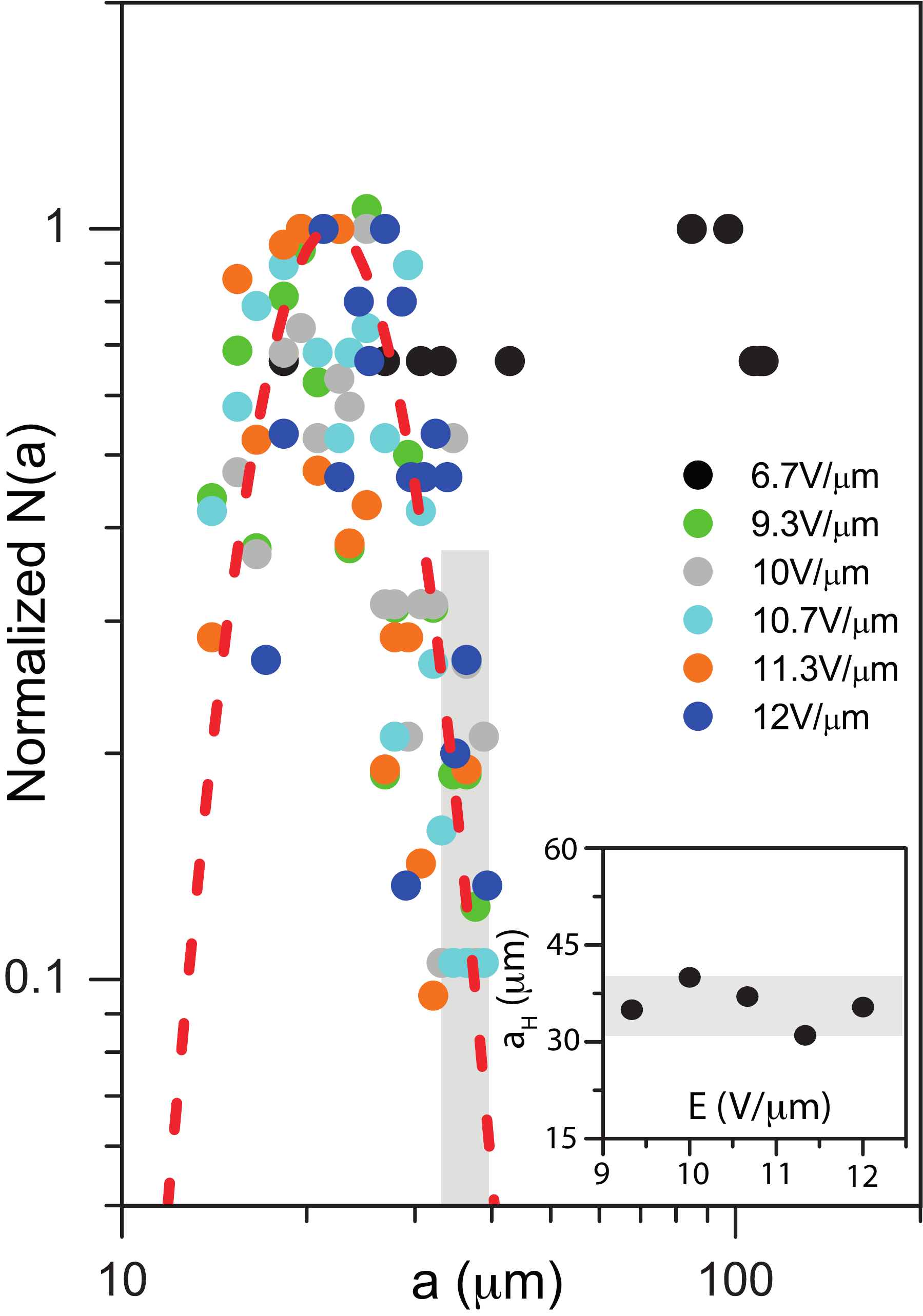}
\caption{The variation of droplet size distribution $N(a)$ normalized with its maximum value with drop radius $a$ for fields above $6~V/\mu m$ is lognormal (red dashed line) with the form: $s(a)=e^{-(log_e(a)-\mu)^2/(2\sigma^2)}/a\sqrt{2\pi\sigma^2}$. The fit yields $log_e$ mean ($\mu)=3.1$ and $log_e$ standard deviation $(\sigma)=0.25$; mean droplet radius $=22\pm 1.3\mu m$, for $E=10~V/\mu m$. The gray bands represent the range of values calculated for $a_H$.  
}
\label{fig5_turb}
\end{figure}
\begin{figure}[htbp]
\includegraphics[scale=0.33]{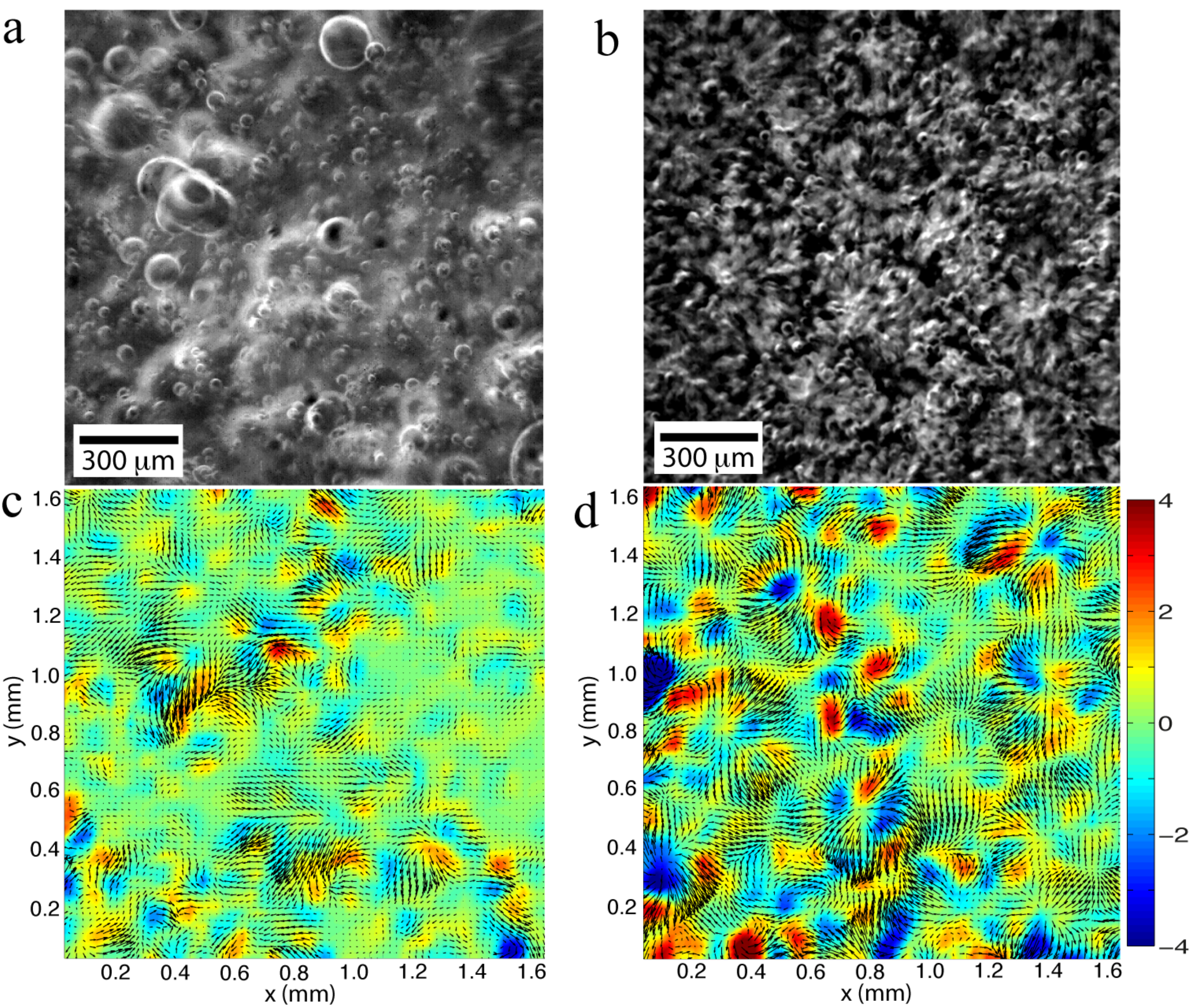}
\caption{Snapshots of the system at (a) $E =  6.7 ~V/\mu m$ and (b) $12~ V/\mu m$. Typical velocity (shown by arrows) and vorticity (the color map in units of $s^{-1}$ ) fields obtained from particle image velocimetry (PIV) for (c) $E = 6.7~ V/\mu m$ and (d) $12~ V/\mu m$, with ${\mathrm v}_{rms}$ = 20 and $100~\mu m/s$ respectively.}
\label{fig1_turb}
\end{figure}

The existence of a typical lengthscale for bubbles in turbulent shear flow was recognized by Hinze \cite{Hinze_1955}, 
who obtained a lengthscale (the Hinze scale $a_H$) by equating the disrupting differential pressure forces on the droplets' surface to the restoring forces due to surface tension,  $a_H \sim \gamma/\rho\langle {\mathrm v}\rangle^2$, where $\rho$ is the density of the liquid phase.  Above the Hinze scale droplet survival is improbable, as reflected in the steady-state droplet size distribution.
The inset of Fig. \ref{fig5_turb} shows the variation of $a_H$  with $E$ (computed, as elswehere \cite{Garrett_2000}, by thresholding the cumulative count of  $N(a)$ at $95 \%$).

The dynamics of the system was simultaneously captured by optical imaging and rheology.
Figure \ref{fig1_turb}(a) and (b) show representative optical micrographs of the system at electric fields $E= 6.7$ and $12~V/\mu m$ respectively.  For $E<6.7 ~V/\mu m$, only a periodic convective motion exists in the system, accompanied by droplet breakup and coalescence events. For increasing values of $E$, however, the flow patterns become random; they initially generate an isotropic chaotic motion of the droplets without producing any distinctive spatial structures ($E= 6.7~V/\mu m$) (see Fig. \ref{fig1_turb}(a) and Supplementary Movie 1).  For $E >10~V/\mu m$,  flow generated coherent structures are observed (see Fig. \ref{fig1_turb}(b) and Supplementary Movie 2). Here, the droplets are not uniformly dispersed; instead, they form dynamic clusters that are reasonably dense (area fractions of  $\sim 0.45$). The corresponding spatial velocity and vorticity maps, shown in Fig. \ref{fig1_turb}(c) and (d) respectively,  are obtained from particle image velocimetry extracted from the motion of droplets. Estimated characteristic velocities ($U_{ev}$) of 20 and 100 $\mu$m/s in Fig. \ref{fig1_turb}(a) and (b) respectively, a channel depth ($d$) of 300 $\mu$m and a diffusion coefficient ($D$) of $2.9 \times 10^{-12} \mathrm{m^2/s}$, the electric Rayleigh number \cite{posner_convective_2006} in our experiment is of order $Ra_E = U_ {ev}d/D \sim 10^3$ to $10^4$. Remarkably, from the same electro-viscous velocities ($U_ {ev}$) and channel depth ($d$), one calculates a Reynolds number $Re  = \rho U_ {ev}d/\eta \sim 3 \times 10^{-3}$ to $10^{-2}$, here $\rho$ and $\eta$ are density and viscosity of castor oil respectively. So this is indeed a system with low $Re$ but high $Ra_E$. 

\begin{figure}[hbt]
\includegraphics[scale=0.33]{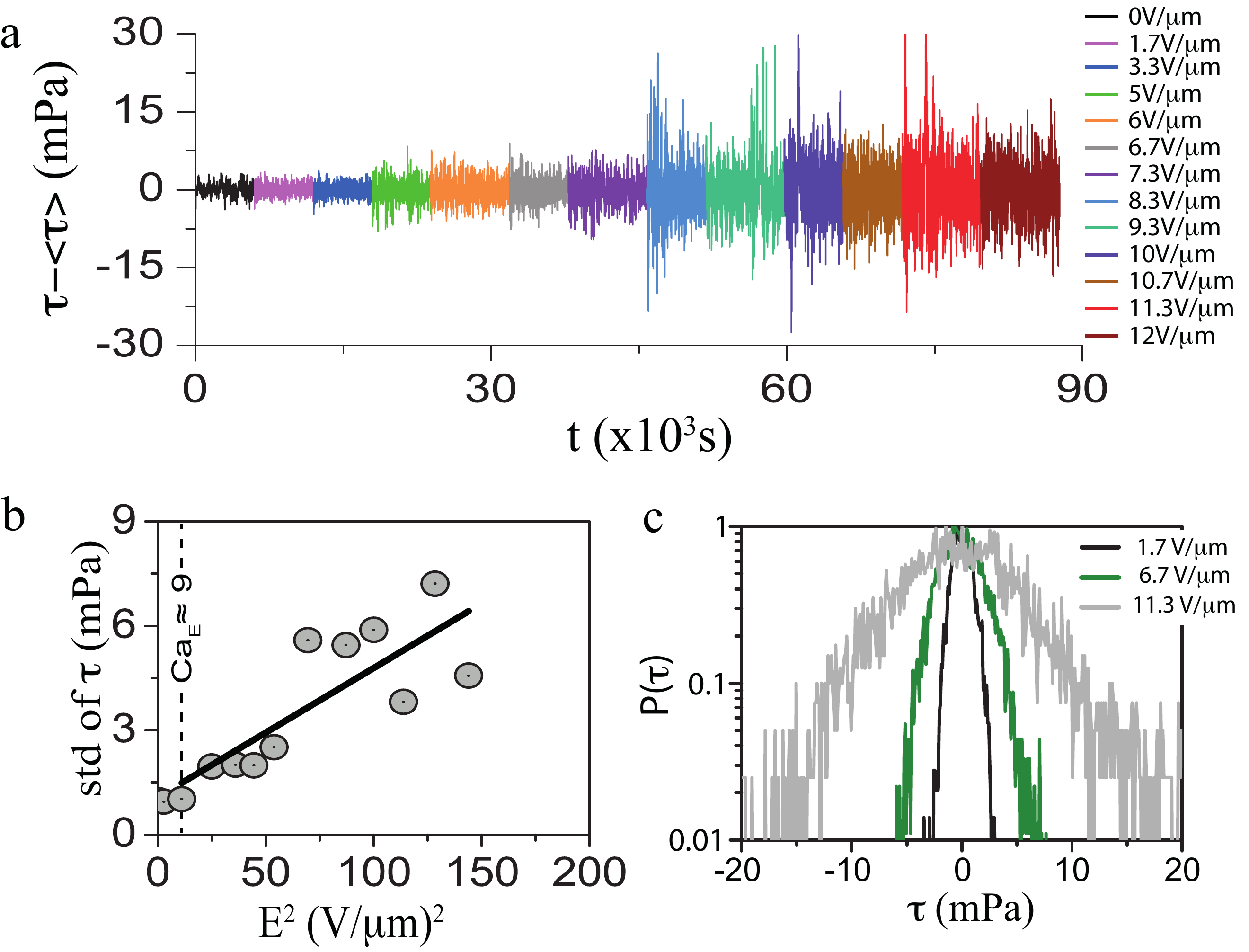}
\caption{(a) Temporal fluctuations of shear stress, $\tau$, with increasing electric field $E$. (b) The standard deviation of shear stress fluctuations plotted against $E^2$ exhibits a linear increase above $Ca_E\approx9$. (c) The normalized probability distribution of ($\tau-\langle\tau\rangle$), $P(\tau)$, broadens with increasing $E$.}
\label{fig2_turb}
\end{figure}

Rheological measurements reveal  time evolution of the electric field induced shear stresses, measured at the top electrode. Figure \ref{fig2_turb}(a) shows the time series of shear stress fluctuations $(\tau-\langle\tau\rangle)$, where $\langle\cdots\rangle$ denotes the time average, for various values (each a different color) of $E$.  We identify it as the Reynolds shear stress, i.e., the contribution of turbulent motion to the mean shear stress \cite{Tennekes_1972}. The data clearly demonstrate  that the internal flow in the cell, which produces fluctuating stress at the walls, become increasingly unsteady with increasing $E$.  In Fig.  \ref{fig2_turb}(b), we observe that the standard deviation of the Reynolds stress, $(\tau-\langle\tau\rangle)$, a measure of the strength of the unsteady flow, grows linearly as a function of the electric field intensity $E^2$ above electric capillary number $Ca_E=\epsilon_0\epsilon_{s}E^2a/\gamma\approx9$, where $\epsilon_0$ is free space permittivity, $\epsilon_s$  is dielectric constant of silicone oil, drop size $a\sim100~\mu m$ and interfacial tension $\gamma=3~mN/m$. 
The corresponding normalized probability distribution of $(\tau-\langle\tau\rangle)$, $P(\tau)$, shown in Fig. \ref{fig2_turb}(c) for $E=1.7$, $6.7$ and
\noindent $11.3~V/\mu m$, closely resemble Gaussian distributions.
The time series of stress fluctuations is averaged over the entire surface of the detector and hence is only sensitive to slowly varying fluctuations happening over a large scale.

The existence of self-organized coherent structures is a characteristic of the inverse energy cascade and is implied as an intrinsic feature of homogeneous turbulent flow \cite{Biferale}. In our system the droplets form cloud-like structures, surrounded by regions sparse in droplets (as shown in Fig. \ref{fig1_turb} with spatially separated regions of high vorticity in red and blue colors). This motivated us to further investigate the properties of field-driven flows  to look for evidence of an unusual energy cascade.

\begin{figure*}[htbp]
\includegraphics[scale=0.40]{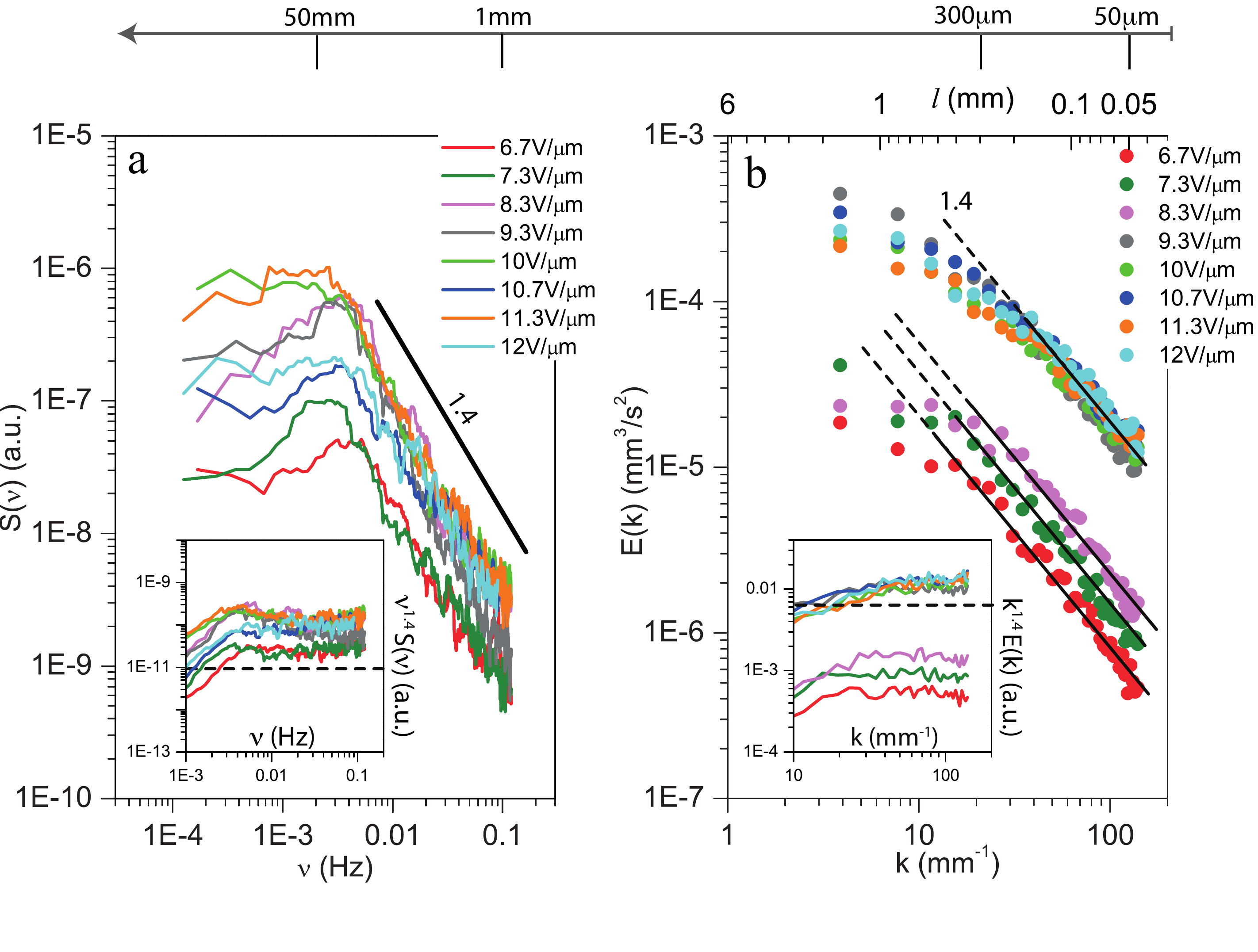}
\caption{(a)  Power spectrum, $S(\nu)$, of shear stress fluctuations obtained at different fields $E$. The power of fluctuations increases
but the frequency dependence of the spectra remains the same in and follows a $\nu^{-\alpha_{\nu}}$ power law dependence for over a decade of frequencies. A global fit for all electric fields to one value of $\alpha_{\nu}$ yields $\alpha_\nu = 1.38 \pm 0.02$, consistent with $\alpha_{\nu} = 1.4$. Inset shows the consolidated plots, i.e., $\nu^{1.4}S(\nu)$, for different electric field values. (b) The energy spectrum, $E(k)$ Vs $k$, of turbulent droplet motion at different field values is globally fit to a $k^{-\alpha_k}$ dependency and yields $\alpha_k = 1.39 \pm 0.02$, once again consistent with a -1.4 power law, shown by the solid black line (the dashed black line indicates the deviation of the results from power law behavior at small $k$). Inset shows the consolidated plots, i.e., $k^{1.4}E(k)$, for different electric field values.}
\label{fig3_turb}
\end{figure*}

Figure \ref{fig3_turb}(a) shows the power spectrum, $S(\nu)$, of the time series data of the Reynold shear stress $(\tau-\langle\tau\rangle)$ for different values of $E$. Strikingly, all the graphs have similar features: there is a peak in $S(\nu)$ at around $2-3~mHz$, and  from $\nu_{low} =3~mHz$ to $\nu_{high} = 0.1~Hz$, the power spectrum exhibits a $S(\nu)\sim\nu^{-\alpha_{\nu}}$ power law. Results at all the fields shown are fit globally to a single value of $\alpha_{\nu}$: this yields $\alpha_{\nu} = 1.38 \pm 0.02$. Assuming a linear dispersion relation $2 \pi \nu \sim {\mathrm v}_{rms}k$, where ${\mathrm v}_{rms}$ is the root-mean-square velocity and $k$ is the wave-vector, the frequency $\nu_{low}$ and $\nu_{high}$ are related to lengthscales $\ell = 2\pi/k = {\mathrm v}_{rms}/\nu $ which are $50~mm$ and $1~mm$ respectively, using ${\mathrm v}_{rms} = 100~\mu m/s$; these lengthscales are thus larger than the cell spacing of $300~\mu m$ but comparable to the lateral extent ($50~mm$) of the system. 

Figure \ref{fig3_turb}(b) shows the energy density $E(k)$ for different values of E, obtained $via$ optical imaging of the droplet motion, coupled with particle imaging velocimetry (see Supplementary Information). It is computed from velocity fields (as shown in Fig. \ref{fig1_turb})
by taking the Fourier transform of the time averaged velocity, i.e., $E(k)$=$F({\mathrm v})F({\mathrm v}^*)/L^2$, where $F({\mathrm v})=FFT ({\mathrm v}-\langle {\mathrm v}\rangle)$ and $L = 1.66~mm$ is the lateral extent of the image. The lowest accessible $k$ is $k_l = 2 \pi/L \sim 3~mm^{-1}$, while the largest, $k_u \sim 100~mm^{-1}$, is set by the distance ($2 \pi/k_u \sim 60~\mu$m) traversed by the droplets between successive images, which is also roughly the same size as the largest droplet diameter. At intermediate values of $k$, we once again observe a power law scaling, $E(k) \sim k^{-\alpha_k}$, with a globally fit value of $\alpha_k = 1.39 \pm 0.02$. Thus, the temporal and spatio-temporal energy spectra have consistent power law scalings, $\alpha_\nu = \alpha_k \sim 1.4$.

What is the nature and underlying mechanism of this turbulent behavior? We do not have a comprehensive answer to this question, but our experiments present two disparate smoking guns.
First, the power laws observed ($\alpha \sim 1.4$) are identical to the ones observed for free 3D Rayleigh-B\'enard (RB) convection in helium gas, at high $Ra_E$ \cite{Wang_2014, Wu_1990, Sano_1989}, and recently in numerical simulations for turbulent flow in RB convection with uniform rotation  \cite{Pharasi_2014}. While electrically-driven RB convection has indeed been observed in leaky dielectrics \cite{Baygents_1998}, an $\alpha \sim 1.4$ power law does not by itself imply RB turbulence.

A second observation points in a different direction. The lower wavenumber of the spectral range of power law scaling varies from $\sim 15$ to $40~mm^{-1}$ as a function of the applied electric field (see Fig. \ref{fig3_turb}(b)). The upper wavenumber, on the other hand, is at $k \sim 150~mm^{-1}$. This obtains a range of lengthscales of 400 to $150~\mu m$, at the upper end, to 40 $\mu$m at the low end. This is remarkable, because the droplet sizes are smaller than this lengthscale. Since the energy injection lengthscales are unambiguously set by the droplet sizes, this implies that the energy is transported towards larger lengthscales, reminiscent of experiments in 2D turbulence \cite{Paret_1997}.

Taken together, the results show an $\alpha \sim 1.4$ power law scaling on lengthscales between $1~mm$ and $50~mm$, and between 40 and 300 $\mu$m (Fig. \ref{fig3_turb}). This is suggestive of an inverse cascade that extends in spatial scales from the droplet scale ($\sim 40~\mu$m) to the lateral extent of the system ($\sim50~mm$), and is an intriguing parallel to the upscale energy cascade seen in a very different system involving atmospheric flows \cite{Xia2011}: in both quasi-2D systems, the ratio of the forcing lengthscale to the system thickness is greater than 0.5, and the ratio of the system thickness to its lateral extent is less than 1\%.

\begin{acknowledgments}
A.Y. thanks members of the TIFR soft matter group for their hospitality during the course of these studies and the Physical Oceanography group at Memorial, and Prasad Perlekar at TIFR Hyderabad, for useful comments. This work was supported in part by NSERC.
\end{acknowledgments}

\bibliographystyle{apsrev4-1}
\bibliography{turbulence_prl}
\end{document}